\documentstyle[12pt]{article}

\newcommand{\eq}{\begin{equation}}
\newcommand{\en}{\end{equation}}
\newcommand{\bea}{\begin{eqnarray}}
\newcommand{\eea}{\end{eqnarray}}
\newcommand{\spz}{\hspace{0.7cm}}

\newcommand{\ba}{\begin{array}}
\newcommand{\ea}{\end{array}}

\newcommand{\resection}[1]{\setcounter{equation}{0}\section{#1}}

\newcommand{\wave}[1]{\mbox{\raisebox{-.6ex}{$\stackrel{\displaystyle{\sim}}
                     {\scriptstyle{{#1}}}$}}}

\newcommand{\iqft}{IQFT$_2$}
\newcommand{\cft}{CFT$_2$}

\newcommand{\G}{{\cal G}}
\newcommand{\h}{{\cal H}}

\newcommand{\NP}[1]{Nucl.\ Phys.\ {\bf #1}}
\newcommand{\PL}[1]{Phys.\ Lett.\ {\bf #1}}

\newcommand{\CMP}[1]{Comm.\ Math.\ Phys.\ {\bf #1}}

\newcommand{\IJMP}[1]{Int.\ J.\ Mod.\ Phys.\ {\bf #1}}

\begin{document}
\sloppy
\renewcommand{\thefootnote}{\fnsymbol{footnote}}

\newpage
\setcounter{page}{0}
\begin{flushright}
Bologna preprint DFUB-93-11\\
October 1993
\end{flushright}
\vskip .3cm

\vskip 1cm
\begin{center}
{\bf INTEGRABLE QFT$_2$ ENCODED\\
     ON PRODUCTS OF DYNKIN DIAGRAMS}\\
\vskip 1.8cm
{\large E.\ Quattrini$^1$, F.\ Ravanini$^1$ and R.\ Tateo$^2$}
     \footnote{E-mail: ravanini@bologna.infn.it, tateo@to.infn.it,
     quattrini@bologna.infn.it}\\
\vskip .7cm
{\em $^1$ I.N.F.N. - Sez. di Bologna, and Dip. di Fisica,\\
     Universit\`a di Bologna, Via Irnerio 46, I-40126 Bologna, Italy\\
\vskip .4cm
     $^2$ Dip. di Fisica Teorica, Universit\`a di Torino\\
     Via P.Giuria 1, I-10125 Torino, Italy}\\
\end{center}
\vskip 1cm

\renewcommand{\thefootnote}{\arabic{footnote}}
\setcounter{footnote}{0}

\begin{abstract}
\noindent
A large class of Thermodynamic Bethe Ansatz equations governing
the Renormalization Group evolution of the Casimir
energy of the vacuum on the cylinder for an integrable two-dimensional field
theory, can often be encoded on a tensor product of two graphs. We demonstrate
here that in this case the two graphs can only be of $ADE$ type. We also give
strong numerical evidence for a new large set of Dilogarithm sum Rules
connected to $ADE\times ADE$ and a simple formula for the ultraviolet
perturbing
operator conformal dimensions only in terms of rank and Coxeter numbers of
$ADE\times ADE$. We conclude with
some remarks on the curious case $ADE\times D$.
\end{abstract}

\vskip.65truein

\resection{Introduction}

The space of two-dimensional Quantum field Theories
is in connection with that of Conformal Field Theories (\cft) in the sense
that the ultraviolet (UV) and infrared (IR) limit of the former must lie in the
space of theories of the latter. In other words, the space of two-dimensional
QFT consists of conformal submanifolds (fixed points) of given central charge
and of Renormalization Group (RG) flows connecting them. Many of these flows
are not integrable, but there exists a subclass of integrable flows that
defines
the space of two-dimensional (euclidean) Integrable Quantum Field Theories
(\iqft).

The challenge to classify all
\iqft\ is one of the most attractive of modern Quantum Field Theory.
However, this task is very far from completion and at the present stage we can
only hope to get partial classifications
similar to what happens in \cft. There also, the whole classification is far
from completed, but for example, the subclass of rational Conformal field
theories (RCFT) is (strongly) conjectured to be organized in series of coset
models~\cite{GKO} $G/H$, thus
putting their description in correspondence
with general properties of the Lie algebras and their affine generalizations.

Analogously in \iqft\ it would be interesting to identify some kind of
general patterns organizing the vaste zoology of integrable models,
in other words some kind of ``Mendeleev table", able to show hierarchical
structures to be later interpreted as effect of some symmetry underlying the
integrable theories. To recognize this organizing criterion can be
useful as the first step to figure out some general property of the yet
misterious symmetries governing \iqft's.

In the present paper we deal with
deformations of some coset CFT $G_k/H_{Ik}$ by one of its relevant scalar
operators, say
$\Phi(x)$ of (left) conformal dimension $\Delta_P < 1$.
Here $G$ is a compact Lie Algebra and $H$ a proper
subalgebra embedded in $G$ with index $I$. The subscripts on $G$ and $H$ denote
the level of the corresponding Kac-Moody algebra.
The theories we are interested in are defined by the action
\eq
{\cal A} = [G_k/H_{Ik}] + \lambda \int d^2x \Phi(x)
\label{one}
\en
where $\lambda$ is a bare perturbing parameter of dimension $y=2(1-\Delta_P)$
and the symbol $[G_k/H_{Ik}]$ stands for the action of the $G_k/H_{Ik}$ coset
CFT.
Such a theory can develop a mass gap and flow to a massive
IR theory as the scale increases, or alternatively keep some of its states
massless and flow towards a non-trivial infrared CFT. Zamolodchikov's
c-theorem~\cite{Zam-cth}
implies, for unitary theories, that the central
charge of the IR CFT must be smaller than that of the UV one.

An \iqft\ possesses by definition an infinite
set of local conserved charges.
This implies~\cite{ZamZam} that the
N-particle S-matrix
is factorizable in terms of 2-particle ones. There are
various ways to conjecture the form of the S-matrix once the conserved local
and non-local charges are known. If the theory is massless, one can circumvent
the problems with the definition of the S-matrix by considering asymptotic left
and right movers, as recently proposed in a series of
papers~\cite{Al4,ZamZam2,FSZ,FS}.

The so far most effective method to recover the UV behaviour of a theory
whose factorizable S-matrix is given is
the so called Thermodynamic Bethe Ansatz (TBA).
That the thermodynamics of a scattering theory
can be reconstructed completely from its
S-matrix was known since the end of the sixties~\cite{BDM}, and the use of
Bethe Ansatz techniques to implement this program for integrable
theories was issued in a seminal paper
by Yang and Yang~\cite{YY} for a non-relativistic scattering problem. More
recently Al.Zamolodchikov has proposed this method to investigate factorized
scattering theories corresponding to IQFT$_2$~\cite{Al1}.

Therefore, any IQFT$_2$ must have a (possibly infinite) set of TBA equations.
It can be seen as a set of equations governing non-perturbatively and exactly
the evolution of the Casimir
vacuum energy of the theory put on a cylinder along the Renormalization Group
flow.
In sect.2 we  present various interesting
systems of TBA equations for some classes
of well known coset CFT perturbed by relevant operators in integrable
directions and
we shall illustrate   how in
most cases the resulting TBA equations can be encoded in a sort of tensor
product of two graphs. This encoding is even more evident by passing from TBA
equations to an equivalent system of functional equations (the so-called
{\em Y-system}) (sect.3).
For the  Y-system to be consistent with TBA interpretation, the two graphs
cannot be chosen arbitrarily. We shall see in sect.3 how one can prove that
they must be both restricted to the set of Dynkin diagrams of ADE type (or
their foldings and their extensions). We also mention two properties of the
$ADE\times ADE$ Y-system, that we checked numerically, namely a set of
Dilogarithm sum rules and a periodicity formula. The amusing fact is that
they can be expressed in terms of properties of Dynkin diagrams only.

The next obvious step is to explore (sect. 4)
the set of would be flows suggested by the
various $ADE$ choices for the two graphs in the product, and recognize them
with known flows or interpret them as signal of new, yet unexplored \iqft's.
We are conscious that the observation of a consistent set of equations does not
prove the effective existence of the flow it seems to describe (although many
reasonable checks like those of sect.4 can be proposed in this direction).
However, it can
be used as a leading tool to investigate, conjecture and then, when possible,
verify the existence of new \iqft's and better understand the relations among
them and the pattern organizing \iqft.
In this respect we trust it as a very productive method.

Of course the set of TBA-like equations we explore here does not exhaust even
the set of \iqft\  of the form (\ref{one}). Our goal is to understand as far as
possible the relation between the $G$ and $H$ algebras in the definition of the
coset UV CFT, and the two Dynkin graphs appearing in the TBA equations, and
possibly find extensions to this set able to incorporate other models.

\resection{TBA equations on $\G\times\h$}

In this section we want to recall some facts known in the literature about TBA
equations and their encoding on products of Dynkin diagrams.

For a clear exposition of the deduction of TBA equations from the diagonal
S-matrix of a purely elastic scattering theory we refer the reader to the
origianl paper of Al.Zamolodchikov~\cite{Al1} and also to ref.~\cite{KM1,KM2}.
In this case, the equations turn out to have the form
\eq
\nu_a(\theta) = \varepsilon_a(\theta)+{1 \over 2 \pi}
\sum_b [\phi_{ab}*\log(1+e^{-\varepsilon_b})](\theta)
\label{two}
\en
where the indices $a,b=1,...r$ label the different species of particles of
masses $m_a$ respectively, the symbol * stands for convolution and the kernel
$\phi_{ab}(\theta)$ is determined from the S-matrix via the formula
\eq
\phi_{ab}(\theta) = -i\frac{d}{d\theta} \log S_{ab}(\theta)
\en
$\theta$ is the rapidity parametrizing the energy (momentum) of a particle as
$m\cosh\theta$ ($m\sinh\theta$). The massive behaviour of the theory is encoded
in the {\em energy terms} $\nu_a(\theta)=m_aR\cosh\theta$ ($R$
being the radius of
the cylinder on which the theory is put, or, by modular invariance, the inverse
temperature). The unknowns of this set of integral equations are the {\em
pseudoenergies} $\varepsilon_a(\theta)$. These in turn determine the Casimir
energy of the vacuum on the cylinder via the equations
\eq
E(R)=-\frac{\pi c(r)}{6R} \spz \mbox{\rm and} \spz
c(r) = \frac{3}{\pi^2} \sum_a
\int_{-\infty}^{+\infty}d\theta \nu_a(\theta) L_a(\theta)
\en
where $L_a(\theta)$ is short for $\log(1+e^{-\varepsilon_a(\theta)})$, and
$r=m_1 R~$ ($m_1$ is the mass of the lightest particle) is such that the UV
limit corresponds to $r\to 0$ and the IR one to $r\to \infty$. The function
$c(r)$ is even more interesting than $E(R)$, as it interpolates directly the
central charges of the UV and IR theories: $c(0)=c_{UV}$, $\lim_{r\to \infty}
c(r)=c_{IR}$ (if the theory is massive $c_{IR}=0$).

Klassen and Melzer~\cite{KM1} explored with this method the whole set of
$\G=A,D,E$ purely elastic minimal S-matrices, confirming
that they describe the set of theories with action
\eq
{\cal A} = \left[ \frac{\G_1 \times \G_1}{\G^{diag}_2}\right]
+ \lambda\int d^2x
\phi^{id,id}_{adj}(x)
\en
where the perturbing operator is here labelled by representations of
$\G\times\G$ (upper indices) and $\G^{diag}$
(lower index) as usual in coset CFT. The link between S-matrix and $ADE$ Lie
algebras can be made explicit in the TBA equations thanks to a transformation
proposed by Al.Zamolodchikov in the very interesting paper ref.~\cite{Al2}. For
all the $ADE$ theories explored in~\cite{KM1} the TBA can be recasted in the
following form
\eq
\nu_a(\theta)=\varepsilon_a(\theta)+{1 \over 2 \pi}
\{ \phi_g*\sum_a \G_{ab}*[\nu_b-\Lambda_b]\}(\theta)
\label{three}
\en
where $\Lambda_b=\log(1+e^{\varepsilon_b})$, $\phi_g(\theta)
=g/2\cosh(g\theta/2)$ is a universal kernel depending on the Coxeter number $g$
of $\G$ only, and all the dependence of the
kernel from the indices $a,b$ has been confined in the incidence matrix of the
Dynkin diagram of the Lie algebra $\G$, that we shall denote in this paper as
$\G_{ab}$. Hence we can think of (\ref{three}) as a set of equations where the
unknowns couple to each other obeying the connectivity of the $\G$ Dynkin
diagram, and in this sense they can be thought as uniquely defined once the
diagram is given. The masses of the particles are known to be encoded on the
Perron-Frobenius eigenvector $ \psi_{\G}$ of $\G$: $m_a=m_1\psi_{\G}^a$.

A large set of \iqft\ is known where the S-matrix is not diagonal. The
deduction
of TBA equations in this case is much more cumbersome, and one has to resort to
Higher Level Bethe Ansatz techniques to diagonalize the "color" transfer matrix
appearing in the Bethe equations. In spite of these difficulties, Al.
Zamolodchikov was able to work out some simple case of TBA
and then to conjecture a generalization valid for the whole set of minimal
models (the $(A_1)_k \times (A_1)_1/(A_1^{diag})_{k+1}$ coset CFT's) perturbed
by their least relevant $\phi_{13}$ operator~\cite{Al3,Al4}.
These equtions are of the form
\eq
\nu_i(\theta) = \varepsilon_i(\theta)+{1 \over 2 \pi}
\sum_j (A_k)_{ij} (\phi_2 *L_j)(\theta)
\label{four}
\en
where the indices $i,j=1,...k$ run on the incidence matrix of the $A_k$ Dynkin
diagram. The most curious feature of this system is the form of the energy
terms. For negative $\lambda$ the behaviour is massive and can be reproduced
by choosing $\nu_i = \delta_i^1 m R \cosh\theta$.
For positive $\lambda$ these theories describe massless
interpolating flows between the $k$-th and $k-1$-th minimal models. In spite of
the conceptual difficulties to define an S-matrix for scattering of massless
objects, one can still present consistent sets of TBA equations with the
concept of particle substituted by that of left and right movers, of energy
$mRe^{\theta}/2$ and $mRe^{-\theta}/2$ respectively. The sets of TBA equations
are basically the same as (\ref{four}), but now $\nu_i=\frac{mR}{2}
(\delta_i^1 e^{\theta} + \delta_i^k e^{-\theta})$.
In both cases, almost all nodes on the diagram are attached vanishing
energy terms. This is a quite general feature of
many \iqft. From the few known examples, one can argue that the
Bethe Ansatz procedure usually performs a sort of
"factorization" of the scattering process in two subprocesses: first a pure
elastic scattering of the
particles as they had no internal indices (colors), then a pure exchange
of color between particles, that one is free to think as mediated by virtual
objects with no energy nor momentum called {\em magnons}, somehow in connection
with the nodes of the diagram having zero energy term, that are therefore
called {\em magnonic nodes}.

Other progresses have been made on the same path. Putting masses on the $k$-th
node or left/right
movers on $k$-th/$l$-th nodes of the $A_{k+l-1}$ diagram, one obtains a TBA set
of equations describing the massive (negative $\lambda$) and massless (positive
$\lambda$) behaviour of all the coset theories $(A_1)_k \times (A_1)
_l/(A_1^{diag})_{k+l}$ perturbed by their $\phi^{1,1}_3$ relevant operator of
dimension $\Delta_P=1-\frac{2}{k+l+1}$~\cite{Al5}

One can also ask if it is possible to use diagrams different from $A_k$ for
this "magnonic" TBA's. In ref.~\cite{FatAl}
the ${\bf Z}_k$ parafermions (in coset
language the $(A_1)_k/U(1)$ RCFT's) deformed by
their $\psi_1\bar\psi_1+c.c.$ operator were studied. The TBA system proposed
there is similar to (\ref{four}) but encoded on $D_k$. See ref.~\cite{FatAl}
for details.
There is an elegant way to write in a unified form all the TBA systems cited
above and also other generalizations. The idea is to introduce ${\bf
two}$ graphs, one encoding the particle (or kink) species ($\G$), the other the
magnonic structure of the colors ($\h$).
The TBA is then fixed by giving the graph tensor product $\G \times \h$ and
specifying the form of the energy term $\nu_a^i(\theta)$ attached to each node
$(a,i)$ (the labels $a,b=1,\dots r=rank(\G)$
run here on the $\G$ graph and $i,j=1,
\dots k=rank(\h)$ run on the $\h$ graph)
\eq
\nu_a^i(\theta) = \varepsilon_a^i(\theta) + \frac{1}{2\pi} \phi_g*
\left\{\sum_b \G_{ab} [\nu_b^i - \Lambda_b^i] - \sum_j \h_{ij}
L_a^j\right\}(\theta)
\label{five}
\en
Here $g$ is simply some real number whose connection with the graph $\G$ shall
specified better below.
The rationale under this generalization comes from the study of
the $\phi^{id,id}_{adj}$ perturbations of
the large class of coset models $\G_k \times \G_l/\G_{k+l}$($
\G=A_n,D_n,E_{6,7,
8}$) , which are the most
starightforward generalizations of the minimal models perturbed by $\phi_{13}$.
They are~\cite{Rav1} the first evident example where one has to introduce
both particle  and magnonic indices. The TBA equations
can be encoded on the {\em product} of diagrams $\G \times A_{k+l-1}$, the
first describing the connectivity of the particle indices, the second that of
the magnonic ones.
The cases previously discussed can all be encoded in this general class
(\ref{five}), with suitable choices of $\G$ and $\h$.
One is naturally lead to ask how large is the scope of integrable flows
described by system (\ref{five}) and if there are restrictions on the choice of
$\G$ and $\h$. We deal with this subject in the next section.

\resection{The Y-system}
Here we wish to recall a very important result
of paper~\cite{Al2} (or better its generalization~\cite{Rav1}
to the more general system (\ref{five})), that perhaps has not
sufficiently been emphasised and appreciated by part of
the S-matrix community, but
that will be the main instrument of our investigation,
i.e. the fact that all solutions of the system (\ref{five}) are also
solutions of the system of functional equations
\eq
Y_a^i\left(\theta-\frac{i\pi}{g}\right)
Y_a^i\left(\theta+\frac{i\pi}{g}\right)=
\prod_b (1+Y_b^i(\theta))^{\G_{ab}} \prod_j (1+Y_a^j(\theta)^{-1})^{-\h_{ij}}
\label{Y}
\en
often referred as the {\em Y-system}. The relation with the TBA equations is
given by $Y_a^i(\theta)=e^{\varepsilon_a^i(\theta)}$.
Here the encoding on the $\G$ and $\h$ graphs is even more evident.
This system has been encountered in many applications of TBA,
 but such kind of objects appear in other areas of physics and
mathematics too, for example in lattice integrable models, $\tau$-functions for
Toda lattices, etc..., so the importance to study such mathematical object
 is larger than the TBA
problem we are considering here.

The derivation from TBA (\ref{five}) given in~\cite{Rav1} is valid for $\G$ any
$ADE$ Dynkin diagram, $g=cox(\G)$ and $\h$ any connected unoriented graph.
One can however consider the
Y-system for any $\G\times \h$ graph and any $g\in R$
and then ask when it is consistent with a
TBA interpretation, in other words when it can be used to represent an
integrable Renormalization Group flow (this does not mean that the flow
{\em exists}, simply that it is  {\em mathematically} possible).
In this connection,
we shall prove the following two basic statements:
\begin{enumerate}
\item
The Y-system (\ref{Y}) can be consistent with a TBA interpretation only if the
graph $\G$ is a Dynkin diagram of a simply-laced simple Lie algebra or a
folding
of it ($ADET$).
This algebra encodes the mass ratios of the particles (or the relative
crossover
scales of the left and right movers) of the integrable QFT underlying it.
\item
if the $\G$ graph is $ADET$ then the $\h$ graph (that is known to
encode the ``color'' structure of the theory) can only be $ADET$ or
at least an extended Dynkin diagram $\hat{A}\hat{D}\hat{E}$ or a folding of it.
\end{enumerate}

\subsection{Proof of statement 1}

The Y-system (\ref{Y}) can be depicted on
the {\em tensor product} diagram $\G \times
\h$, organized in {\em $\G$-rows} reproducing the $\G$ diagram, and
{\em $\h$-columns} reproducing the $\h$ diagram. We wish to prove
that a physically relevant system of this kind can only exist for $\G$
in the set of simply-laced Dynkin diagrams or foldings of them $ADET$.
{\em Physically relevant} means that we
are interested in systems of the kind (\ref{Y}) that allow an
interpretation in terms of a scattering theory; i.e. they must come from
some TBA, able to reproduce sensible physical behaviours at UV and IR.
As $Y_a^i(\theta) = e^{\varepsilon_a^i(\theta)}$ and
the asymptotic behaviours for large $R$ and large $\theta$ of the
pseudoenergies are driven by the corresponding energy terms
$\nu_a^i(\theta)$, the possible behaviours of $Y_a^i(\theta)$ are of the
following two types:
\eq
Y_a^i(\theta) \wave{R,\theta\to +\infty} \left\{
\begin{array}{ll}
e^{m_a^i R e^\theta/2} & \mbox{if the node $(a,i)$ is massive or left mover}\\
e^{const.} & \mbox{if the node $(a,i)$ is magnonic or right mover}
\end{array} \right.
\en
In what follows we denote the nodes having the massive or left mover
behaviour as {\em black}; those with magnonic or right mover behaviour
as {\em white}.

If a system of type (\ref{Y}) has white nodes only, i.e. if the
corresponding TBA has all the $\nu_a^i(\theta)\equiv 0$, it degenerates
completely to have the totally trivial solution $c(r)\equiv 0$.
Therefore a physically sensible
Y-system must have at least one black node. We now prove that if a
$\G$-row contains a black node, then all the nodes in that row must be
black. Indeed, consider a would-be white node $a$ on the same $\G$-row,
connected to the black node $b$. Connected means that $\G_{ab}
\not= 0$. On the left hand side of (\ref{Y}) we then have a constant
asymptotic behaviour. On the right hand side the product over $\h$ also
behaves as a constant irrespective of the color of the nodes adjacent to
$a$ in the $\h$-column direction. However, the product over $\G$
contains at least one term (the one connecting $a$ with $b$) that has
the exponential behaviour typical of black nodes. The constant behaviour
is exponentially depressed compared to the ``black'' one, therefore, comparing
terms of order $e^{\theta}$ in the asymptotic of (\ref{Y}) we would get the
condition
\eq
0=m_b^i R e^{\theta}/2
\en
which would imply that the $b$ node has zero mass, in contradicion with
the hypotesis that it is black. Hence we conclude that for a given
$\G$-row the nodes are all black or all white.

We said that the Y-system must have at least one black node: this
implies that indeed there must be at least one black $\G$-row. With this
result at hand, we proceed further by picking up a black $\G$-row for a
fixed $\h$ index $i$, and considering the asymptotics of system
(\ref{Y}) for this row. As before, the $\h$ product on the r.h.s. drops
having constant behaviour, while the $e^{\theta}/2$ terms give the
asymptotic consistency condition
\eq
2 m_a^i \cos \frac{\pi}{g} = \sum_{b=1}^{r} \G_{ab} m_b^i
\label{PF}
\en
{}From this equation some very important facts follow:
\begin{enumerate}
\item
if $\G=A_1$, i.e. the $1\times 1$ matrix 0, then the r.h.s. of (\ref{PF})
vanishes. For the l.h.s to vanish too, we must have $g=2$.
\item
if $\G$ is the incidence matrix of a non-trivial connected graph, then the
r.h.s. of (\ref{PF}) is positive, therefore, to have a positive l.h.s. we must
require $g>2$.
\item
for $2<g<\infty$, as $\cos\frac{\pi}{g}<1$ we have the condition
\eq
\sum_{b\in\G} \G_{ab} m_b^i < 2 m_a^i
\label{Dynkin}
\en
where the vector $m^i=(m_1^i,m_2^i,...,m_{r}^i)$ has all positive
components. This is exactly the condition that selects, among all
possible connected graphs the list of simply-laced Dynkin
diagrams and their possible foldings,
\eq
\G ~=~ A_n~,~D_n~,~E_{6,7,8}~,~T_n=A_{2n}/Z_2
\en
\item
As the vector $m^i$ is an eigenvector of $\G$ with all non negative
components, it must be proportional to the (unique) Perron-Frobenius
eigenvector $\psi^{(\G)}$ of $\G$
\eq
m_a^i = m^i \psi_a^{(\G)}
\en
for each black $\G$-row. Moreover, $g$ plays the role of the (dual) Coxeter
number of $\G$, hence it is a positive integer.
\end{enumerate}
\subsection{Proof of statement 2}
Now we prove another statement, i.e. if $\G$ is a simply-laced
($ADET$) Dynkin diagram then
$\h$ is or in the set $ADET$ too, or in the set of extended simply-laced Dynkin
diagrams or their foldings. This can be done by generalizing an argument given
in~\cite{RTV}. Consider stationary solutions (i.e. indipendent on
$\theta$) of the system (\ref{Y}). We know that these solutions enter
the calculation of the UV central charge and in particular that at least
one solution with real $\varepsilon_a^i$
must occur (for the simplest cases one can prove that
this real solution exists and is unique, we shall assume this as true
for all cases in what follows). Reality and finiteness
of the $\varepsilon_a^i$ implies that $y_a^i>0$. Then
\eq
2 \log y_a^i = \sum_b \G_{ab} \log (1+y_b^i) - \sum_j
\h_{ij}\log(1+1/y_a^j)
\en
The quantity $z_a^i=\log(1+y_a^i)$ is real and strictly positive, which
implies that it satisfies $\sum_b \G_{ab}z_b^i < 2z_a^i$, as $\G$ is
in the list of $ADET$ Dynkin diagrams. This allows,
after few manipulations, to obtain the inequality
\eq
\sum_j\h_{ij}w_a^j < 2w_a^i
\label{H}
\en
where $w_a^i=\log(1+1/y_a^i)>0$. This proves that $\h$ must also be in
the list of $ADET$ Dynkin diagrams.
It can happen that the system degenerates to have some $y_i^a=0$ i.e.
$\epsilon_i^a=-\infty$.
In this case one can prove, using tecniques similar to those of the proof of
statement 1, that if one node has $y_i^a=0$, then all nodes have $y_i^a=0$.
In this case instead of (\ref{H}) we get the equality
$\sum_j\h_{ij}w_a^j = 2w_a^i$ selecting the extended Dynkin diagram
$\hat{A}_n,\hat{D}_n,\hat{E}_{6,7,8}$ or the vaste set of their foldings.
We have to mention here also this somehow degenerate case, as it has physical
applications, e.g. in~\cite{sausage}

\subsection{Some useful formulas on the Y-system}

Many authors converge on the opinion that the Y-system seems to encode a
great amount of information about the \iqft\ it is attached to. In particular
the connection between the Y-system and the Rogers Dilogarithm function seems
to be very productive in reproducing the properties of the UV and IR limits of
the integrable model under consideration~\cite{Nahm}.
One can extremize this point and think to a {\em
reconstruction program}, i.e. to an answer to the question: given a Y-system
and the asymptotic behaviour of its solutions for $\theta,R\to +\infty$,
can we reconstruct completely an \iqft\ underlying it? Is this theory unique?
This program is up to now very far from completion. However many facts indicate
that it is not hopeless. We shall not deal with this very interesting aspect in
this paper; we shall be content to comment here about the two simplest
quantities that can be extracted
from a Y-system and their relation to the graphs $\G \times \h$.
\begin{enumerate}
\item
The stationary solutions $Y_a^i$
of the Y-system allow to compute
the central charge via Dilogarithm sum rules. The main ingredient is the sum
\eq
s(\G\times\h) = \sum_{a\in\G} \sum_{i\in\h}
{\cal L}\left(\frac{1}{1+Y_a^i}\right) =
\frac{\pi^2}{6} \frac{rk}{g+h} h
\label{sumrule}
\en
Here ${\cal L}(z)$ denotes the Rogers Dilogarithm function, $g=cox(\G)$ and
$h=cox(\h)$.
For $\G=A_r$ and $\h=A_k$ this sum rule is proven (\cite{kir}), we checked it
numerically to very high precision up to $r=rank(\G)$ and $k=rank(\h)$ both
equal to 50 for the remaining cases, for which
this sum rule is new and is proposed to the
mathematicians for a rigorous proof.
\item
The solutions of the Y-system
possess a periodicity $Y_a^i(\theta+i\pi P)=Y_a^i(\theta)$ ($P\in{\bf Z}/g)$)
In~\cite{Al2,Al4} arguments are given to relate $P$ to the
conformal dimension $\Delta_P=1-1/P$ of the UV perturbing operator
and that of the IR attracting operator if it is the case. We checked
numerically the following formula for the periodicity to be valid for all the
$\G\times\h$ Y-sytems
\eq
P=\frac{g+h}{g}
\label{periodicity}
\en
\end{enumerate}

\resection{Zoology of $\G\times \h$ \iqft}

Armed with the properties and formulas of the preceeding section, one can
explore the whole set of $ADET \times ADET$
cases. Here we only briefly report on a very preliminary exploration where we
only computed UV and IR central charges, thanks to (\ref{sumrule}) and
conformal dimensions of UV perturbing and IR attracting operators, thanks to
(\ref{periodicity}). The standard way to extract these data from the $\G\times
\h$ Y-system (or TBA) is widely explained in the literature, see
e.g.~\cite{KM1,Al4}. In the following we just comment on some interesting
cases. Here $\G=ADE$.
\begin{itemize}
\item $\G\times A_{k+l-1}$ is the case discussed in~\cite{Rav1}. It corresponds
to the best studied class of coset models, namely $\G_k\times\G_l/G_{k+l}$,
perturbed by the operator $\phi^{id,id}_{adj}$. We refer to~\cite{Rav1} for the
details.
\item A complete list of all the $A_1\times ADET$ cases is given in~\cite{RTV}.
\item A set of interesting cases where one is able to reconstruct the S-matrix,
that turns out to be diagonal, is $\G\times T_1$~\cite{RTV2}.
All the cases involving the
$T_n$ diagrams correspond to non-unitary theories. There are curious
developments in this direction~\cite{RST}, in particular for $T_1\times A_k$.
\item A partially new and interesting case is $\G\times D_k$ with a massive
energy term on one node of the fork, or left and right movers on the two nodes
of the fork. This case generalizes to higher $\G$ what has been discussed
in~\cite{FatAl} for $A_1\times D_k$. In that paper this TBA has been used to
describe $Z_k$ parafermionic theories perturbed by their $\psi\bar{\psi} +
\psi^{\dagger}\bar{\psi}^{\dagger}$ operators ($\psi$ being the generating
parafermion). In the massless direction these theories flow to an IR limit
given by the minimal models $(A_1)_k\times (A_1)_1/(A_1)_{k+1}$. Taking the
limit for large $k$ one defines a flow from $c=2$ to $c=1$ which is interpreted
as the
$O(3)$ massless sigma model with theta term $\theta=\pi$, which
renormalizes at IR to the $SU(2)_{k=1}$ WZW model. One would search for a
generalization of this class of flows to higher $\G$. In a certain sense the
most obvious candidate is exactly this set $\G\times D_k$. With the choice of
left and right movers on the two nodes of the fork, this TBA does indeed
reproduce an IR limit towards $\G_k\times\G_1/\G_{k+1}$. The UV CFTs, however,
are not the $\G_k/U(1)^r$ as one would naively expect by generalizing the $A_1$
case, but better the list of so called {\em Normal Forms} over $\G$ (for a
definition and a list, see the appendix in~\cite{Per1}).
This suggests that the corresponding sigma models
defined as the $k\to\infty$ limit of these series are {\em quantum} integrable
and that it is possible to add some kind of topological term transforming the
model in a massless one with IR limit the $\G_{k=1}$ WZW model. However,
it seems that these sigma models (apart the case $A_1/U(1)$) do not
allow topological terms~\cite{Per2} and, to our knowledge, even the quantum
integrability is an unclear issue here. Moreover, another fact seems to
invite to deepen the study of these models. In the $A_1$ case, it has recently
proposed a so called {\em sausage model}, i.e. a deformation of the target
space
of the $O(3)$ sigma
model defining a one parameter family of integrable sigma models.
A TBA has been proposed for a discrete set of values of this parameter, encoded
on $A_1\times \hat{D}_k$. The main feature is that all the UV central charges
of these TBA's are equal to 1, irrespective of the rank of the $D$ diagram,
thus allowing the sausage interpretation. Now, if this interpretation can be
extended to higher $\G$, one should expect constant central charges, usually
equal to some integer. This is not the case for all $\G\times \hat{D}_k$ with
$\G$ other than $A_1$. The
generalized sausage models, if they exist, have nothing to do with the sigma
models on the normal forms of $\G$. We intend to return on this puzzling
problem in the future.
\end{itemize}

To conclude, we have explored here a wide class of sets of TBA equations, that
incorporate almost all examples shown so far. We have determined definitively
the scope of this class, and, exploring the zoology of the models encoded in
it, we have found repeated regularities but also some unclear problems to be
investigated further. Of course, we are conscious that this class of \iqft's
does not exahust all the possible deformations of RCFT's. A well known
counterexample is the set of $\G_k\times\G_l/\G_{k+l}$ CFT deformed by their
$\phi^{id,id}_{adj}$ operator, where $\G$ is a non-simply-laced algebra. This
case needs a TBA, and then a Y-system, which is not included in the set
described above. The most reasonable generalizations one can think are
\begin{itemize}
\item to allow {\em different} diagrams $\h_a$ on different nodes of $\G$
\item to allow for shift terms on the right-hand-side of the Y-system too.
\end{itemize}
This is indeed the case for the non-simply-laced $\G\times\G/\G$. We shall
report on this in a forthcoming publication.

{\bf Acknowledgements} -- We are grateful to P.Dorey, P.Fr\`e
F.Gliozzi, A.Kirillov, G.Sotkov, M.Stanishkov and
A.Valleriani for useful discussions and help.
R.T. thanks the Theory Group at Bologna University for the
kind hospitality during various stages of this work.

\end{document}